# Topological photonics in one-dimensional settings

Shiqi Xia[1,+], Ziteng Wang[1,+], Domenico Bongiovanni[1,2,+], Dario Jukić[3],
Daohong Song[1,4], Liqin Tang[1,4], Jingjun Xu[1], Roberto Morandotti[2], Hrvoje Buljan[1,5], and Zhigang Chen[1,4]

[1] The MOE Key Laboratory of Weak-Light Nonlinear Photonics, TEDA Applied Physics Institute and School of Physics, Nankai University, Tianjin 300457, China
[2] INRS-EMT, 1650 Blvd. Lionel-Boulet, Varennes, Quebec J3X 1S2, Canada
[3] Faculty of Civil Engineering, University of Zagreb, Zagreb, Croatia
[4] Collaborative Innovation Center of Extreme Optics, Shanxi University, Taiyuan, Shanxi 030006, People's Republic of China
[5] Department of Physics, Faculty of Science, University of Zagreb, Bijenička c. 32, 10000 Zagreb, Croatia

[+] These authors contributed equally

**Abstract:**

Over the past decade, topological photonics has emerged as a vibrant field, attracting significant attention and witnessing remarkable advancements. This growth can be attributed to its fundamental appeal and the unique opportunities it offers for unconventional control of light, promising innovations in next-generation photonic devices. At the heart of topological photonics lies the one-dimensional (1D) Su–Schrieffer–Heeger (SSH) model. Originally conceived to elucidate the physics of a molecular chain of polyacetylene, this model has found widespread applications in exploring a wide range of topological phenomena in photonics and beyond. In this chapter, we aim to provide an overview of topological photonics in one-dimensional (1D) settings. After briefly introducing paradigmatic 1D models, including the SSH, Rice–Mele, and Aubry–André–Harper (AAH) models, we review recent advances in experimental studies and applications of topological photonics based on 1D platforms. Our discussion highlights demonstrated examples, such as the nonlinear tuning of topological states in both Hermitian and non-Hermitian photonic SSH lattices, as well as nonlinear harmonic generation and topological lasing in SSH-type photonic microstructures. We further discuss characteristic topological phenomena in other representative 1D settings, including Floquet systems, topological pumping, quasicrystals, and synthetic non-Hermitian systems. Finally, we examine selected examples of two-dimensional (2D) photonic topological crystalline insulators that are closely linked to the SSH model. Towards the end, we summarize the chapter and provide a list of key contributions, together with an outlook on possible future directions in 1D topological photonics. While this review focuses specifically on 1D topological photonics, it is not intended to be comprehensive or exhaustive.

# I. Introduction

Topological photonics has emerged as a dynamic and rapidly evolving field, drawing inspiration from mathematical topology and early investigations into quantum Hall effects and topological insulators in condensed matter physics [1-5]. The theoretical groundwork for optical analogs of quantum Hall edge states was laid by Haldane and Raghu in 2008, utilizing photonic crystals with broken time-reversal symmetry [6]. However, the experimental realization of these concepts encountered challenges in optics, particularly due to the weak magnetic field response of optical materials, leading to initial demonstrations in the microwave domain [7]. Subsequently, groundbreaking experimental demonstrations and theoretical proposals unveiled various forms of photonic topological insulators [8-15], spanning Floquet photonic topological insulators, aperiodic coupled resonators, topological metamaterials, and crystalline topological insulators. These pioneering efforts rapidly propelled topological photonics to the forefront of interdisciplinary sciences [14-23].

In one-dimensional (1D) systems, the Su-Schrieffer-Heeger (SSH) model stands out as one of the prominent and simplest topological models [24]. Renowned for its chiral symmetry and the emergence of topological edge modes under nontrivial Zak phase conditions [25], the SSH model has been instrumental in numerous experimental realizations and theoretical explorations in topological photonics. Experimental demonstrations of optical Shockley-like surface states in photonic SSH lattices exemplify the existence of these topologically nontrivial edge modes (Fig. 1a) [26]. Introduction of optical nonlinearity in such SSH lattices has unveiled new properties, enabling control of topological zero modes, spectral tuning, and phase transitions. The interplay of nonlinearity and topology has led to both inherited and emergent topological phenomena [27-30]. By juxtaposing two SSH chains with different topological phases, the formation of topologically protected interface states can be realized, with the interface acting as a topological "defect" [31]. Such topological defects in interfacing SSH lattices have served as a platform for studying a diverse range of phenomena, from topologically protected quantum states (Fig. 1b) [32] to dissipative gap solitons (Fig. 1c) [33], nonlinear parametric amplification (Fig. 1d) [34], and topological lasing in hybrid silicon microring resonators (Fig. 1e) [35]. Additionally, topological funneling of light due to the non-Hermitian skin effect has been observed (Fig. 1f) [36].

In recent years, the scope of investigations within the SSH model has expanded to encompass the interplay of nonlinearity, topology, and non-Hermiticity [37]. Furthermore, studies have explored topics such as topological phase extraction (Fig. 1f) [38] and the realization of fully controllable topological networks (Fig. 1g) [39], particularly in the context of synthetic dimensions [40-42].

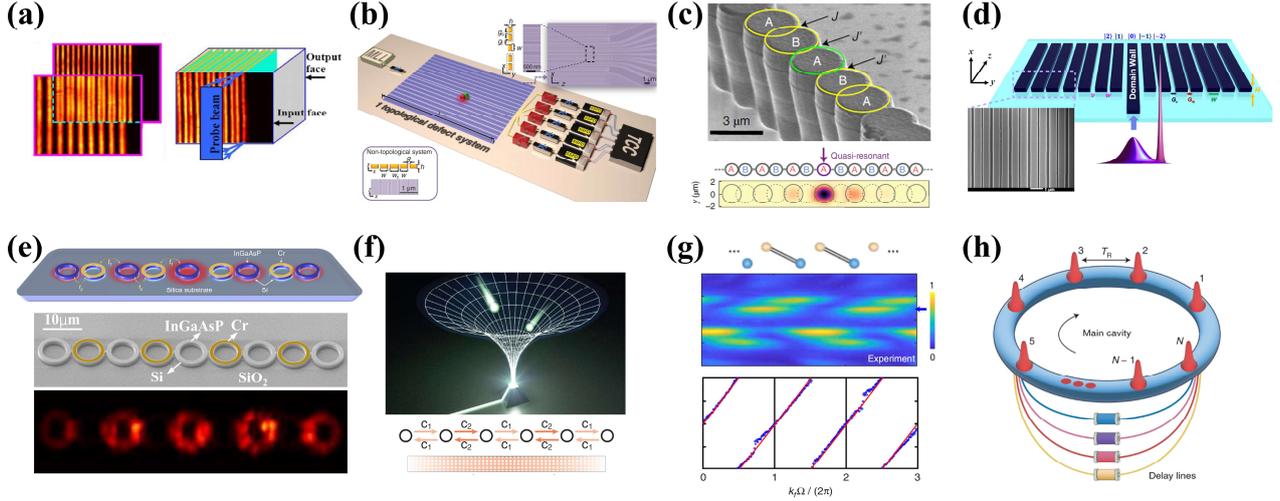

**Fig. 1 Early realizations and recent advances of the SSH lattices in photonics.** (a) Early experimental realization of the SSH lattice and edge states in optics [26]. (b) Experimental demonstration of topologically protected biphoton states using an SSH lattice [32]. (c) Gap solitons in a driven-dissipative SSH polariton lattice [33]. (d) Nonlinear parametric amplification in the SSH model with a domain wall [34]. (e) Topological micro-lasers realized in an SSH-based hybrid silicon array [35]. (f) Topological funneling of light realized in a non-Hermitian SSH lattice [36]. (g) Extraction of topological Zak phase from the SSH-based bulk band structures in synthetic frequency dimensions [38]. (h) A time-multiplexed photonic resonator network for the realization of a dissipatively coupled SSH model [39]. These panels illustrate the diverse range of phenomena and applications associated with SSH lattices in photonics, showcasing their versatility and potential in various optical systems and devices.

In this chapter, we provide an overview of recent advancements in 1D topological photonics, with a specific focus on photonic SSH lattices for realizing and controlling both linear and nonlinear topological edge states. We begin in Section II with a brief introduction to the theory of the 1D topological SSH model, highlighting its significance and connections to other 1D models that may relate to the 2D quantum Hall effect. Section III presents a concise review of recent experimental progress, detailing our methods for generating 1D photonic lattices in photorefractive nonlinear crystals. We also discuss recent advancements in nonlinear topological photonics achieved through optically induced photonic lattices. In Section IV, we explore recent developments in topological photonics within various other 1D settings beyond the SSH model, broadening our understanding of topological phenomena in different contexts. Moving to Section V, we briefly review SSH-type 2D systems aimed at establishing a bridge from 1D to higher-order topological insulators, offering insights into the extension of topological concepts across dimensions. Finally, in Section VI, we address challenges and outline future directions, particularly focusing on the prospects and challenges of topological photonics in 1D settings.

Over the past decade, topological photonics has undergone rapid development, marked by sustained research interest and significant discoveries. While we incorporate selected recent progress on one-dimensional topological photonic systems from the past two years toward the end of this review, as with other reviews in the field [14-23], the present chapter has a focused scope and does not aim to be exhaustive.

## II Topological models in 1D systems - Theory

In this section, we provide an overview of the theoretical development of 1D topological physics, focusing on several concrete systems where topological concepts are investigated. We start by introducing the paradigmatic SSH model, which serves as a fundamental example of 1D topological insulators. Through this model, we illustrate key concepts such as topological invariants, bulk–boundary correspondence, and symmetry-protected topological phases. Following that, we briefly discuss the Rice-Mele model and the off-diagonal Aubry−André−Harper (AAH) model. These models serve as examples to illustrate concepts like topological adiabatic pumping and the mapping from 1D systems to the 2D quantum Hall effect [43-45]. By exploring these concrete systems and theoretical frameworks, we aim to introduce the rich landscape of 1D topological physics and its implications for understanding topological phenomena in photonics as well as condensed matter systems.

**2.1 Fundamental topological concepts regarding the SSH model**

The SSH model depicts spinless particle hopping (coupling) on a 1D dimerized chain with different intercell and intracell hopping amplitudes. As illustrated in Fig. 2a, the chain consists of $N$ unit cells, each of which hosts two sublattice sites ($A$ and $B$) with equal onsite energy (taking, for example, zero for simplicity), and $v$ and $w$ represent the intracell and intercell hopping amplitudes. Under the tight-binding conditions, the SSH Hamiltonian takes the form

$$H = \sum_N v a_n^\dagger b_n + w a_{n+1} b_n^\dagger + h.c., \qquad (1)$$

where $a_n^\dagger(b_n^\dagger)$ are the creation operators for $A(B)$ sublattice sites on the $n$-th unit cell. In general, topological phase typically requires a specific symmetry for protection. Here, for the 1D SSH system, it is naturally guaranteed by chiral symmetry or inversion symmetry.

In the context of single-particle tight-binding models, chiral symmetry simply implies: for a given lattice that can be divided into two sublattices ($A$ and $B$), the hopping only occurs between $A$ and $B$ sublattices, i.e., there is no hopping between the same kind of sublattices. This property leads to an off-diagonal Hamiltonian, typically expressed in $k$-space as

$$H(k) = \begin{pmatrix} 0 & h^\dagger(k) \\ h(k) & 0 \end{pmatrix}. \tag{2}$$

A topological invariant is subsequently defined by $h(k)$, reads [46]

$$\mathcal{W} = \frac{1}{2\pi i} \int_0^{2\pi} dk \frac{d}{dk} \log[\det h(k)]. \tag{3}$$

In a geometrical understanding, $\mathcal{W}$ indicates how many times $\det h(k)$ winds around the origin or a reference point as $k$ is varied by $2\pi$, thus giving rise to the definition of $\mathcal{W}$ as "winding number". The winding number always takes integer values and remains unaffected by gap-preserving perturbations. For a finite-sized SSH chain with the open boundary condition, the value of the winding number corresponds to the number of topological edge states that exist in the system. Such edge states are bounded at the end of the chain and always reside at zero-energy. Therefore, they are also regarded as "topological zero-modes". Specifically, for the SSH lattice, we have $\mathcal{W} = 1$ for $v < w$, and $\mathcal{W} = 0$ for $v > w$. For the former condition, the SSH lattice is in a topologically non-trivial regime and hosts zero-energy edge states [47], whereas for the latter case, it is instead in a topologically trivial regime and thus does not support any topological edge states.

The correspondence between the winding number and the number of topological edge states is rigorous and has been established as the standard "bulk-boundary correspondence," which has been successfully applied to describe diverse classes of topological phases [48, 49]. In the context of the SSH model, it is valid only in the presence of chiral symmetry, and its topological phase is therefore regarded as a "symmetry-protected topological (SPT) phase". The supported edge states are robust (pinned at zero-energy) under the chiral-symmetry preserving perturbations. It is worth mentioning that, in real practical implementations, chiral symmetry cannot be rigorously maintained due to realistic imperfections and constraints. Consequently, a realistic edge state may deviate from the perfect zero-energy mode. However, as long as the edge state resides close to the mid-gap, it inherits topological features as a nontrivial edge-localized state, as demonstrated in many experiments.

Chiral symmetry can be realized in versatile photonic platforms, including waveguide arrays, micro-ring resonators, and photonic crystal nanocavity arrays [14, 18, 50]. In these platforms, the winding number can be appropriately used as a topological invariant. Nevertheless, there are platforms where chiral symmetry cannot be readily satisfied, e.g., the conventional photonic crystals [51]. In such scenarios, the winding number is no longer a good topological invariant.

If a system respects the inversion symmetry, the Zak phase can be exploited as an appropriate choice for the topological invariant [25], defined as

$$\mathcal{Z} = -\int_0^{2\pi} i dk \langle \psi_j | \partial_k | \psi_j \rangle, \tag{4}$$

where $\psi_j$ is the Bloch wave function of the $j$-th energy band. $\mathcal{Z}$ (mod $2\pi$) is quantized to $\pi$ or $0$, respectively corresponding to the topologically non-trivial or trivial regime. $\mathcal{Z}$ acquires an explicit physical meaning with respect to the modern polarization theory in terms of the Berry phase [52]. Briefly speaking, the bulk polarization, defined as $P = \mathcal{Z}/2\pi$, gives quantized positions where the electrons are occupied (Wannier centers) within a unit cell. If $\mathcal{Z} = \pi$ ($P = 1/2$), the Wannier centers are located at the boundary of each unit cell (Fig. 2b1), resulting in a $1/2$ quantized fractional charge [53, 54]. On the contrary, when $\mathcal{Z} = 0$ ($P = 0$), the Wannier centers are located at the middle of the unit cell, implying a topologically trivial regime with zero fractional charge (Fig. 2b2).

The investigation of the SSH model and its underlying physics has a long and venerable history. Early in 1939, Shockley proposed an $sp$ hybridized model and used it to study surface states due to band crossing in a 1D *periodic potential* field [55], which possesses the same physical picture as the SSH model [24]. Later in 1976 [56], Jackiw and Rebbi discovered that the existence of the 1D zero mode is rigorously connected with the structure of the Dirac mass, which is commonly referred to as the "Jackiw-Rebbi" model for studies of soliton-monopole systems in high-energy physics [56]. Intuitively, the SSH model can be equivalent to the Jackiw-Rebbi mechanism: at low-energy approximation, it is governed by Dirac equations, and the Dirac mass term reads $M = v - w$. When a domain wall is constructed with two SSH chains (Fig. 2c), we have a negative Dirac mass term on the right side, but a positive Dirac mass on the left side. Since the sign of the Dirac mass changes at the domain wall, there must be a bounded zero mode at the interface. Extending the concept into the 2D domain, the well-known Jackiw-Rossi model was proposed in 1981 for studies of zero modes of the vortex-fermion systems [57]. Indeed, the equivalence between the SSH model, the Shockley model, and the Jackiw-Rebbi model has been well established. Decades later, the observation of optical Shockley-like surface states [26] was recognized as the first experimental realization of the SSH model and its associated topological edge states in photonics [8, 14, 16]. Importantly, the equivalence and underlying physics of the 1D topological models were later discovered to be responsible for the standard vertical-cavity surface-emitting laser technologies, leading to the proposal of a 2D Dirac-vortex cavity for experimental realization of topological-cavity surface-emitting lasers [58, 59].

Although the theory regarding the standard SSH model was already formulated long ago, there have been constant efforts on new studies and applications of the model recently, as the field of topological photonics has grown dramatically. For example, in 2021, Jiao et al. demonstrated an extended SSH model that breaks the conventional bulk-boundary correspondence [60]. In 2023, Wang et al. proposed and demonstrated a new concept of sub-symmetry-protected topological phase using a perturbed SSH model, which greatly extends the understanding and common scope of topological physics [61]. With the interplay of optical nonlinearity [28-30], quantum and nanophotonics [31, 32], and non-Hermitian topological photonics [28, 62], intriguing phenomena and rich physics from even the simplest 1D topological model are yet to be fully explored.

## 2.2 Topological adiabatic pumping

The concept of "topological pumping" in 1D systems typically arises from an effective mapping between a 1D topological model and a 2D topological insulator with a non-zero Chern number (e.g., the integer quantum Hall effect). We first introduce this concept through the Rice-Mele model, which is simply a modified SSH model with additional staggered onsite potential $u$, captured by the Hamiltonian

$$H = \sum_N v a_n^\dagger b_n + w a_{n+1} b_n^\dagger + u(a_n^\dagger a_n - b_n^\dagger b_n). \tag{5}$$

In the context of topological pumping, $u, v,$ and $w$ are assumed to be time-dependent (in some experimental platforms, such as waveguide arrays, the propagation length $Z$ is treated equivalently as the time). For example, one appropriate setting is: $u = \sin\left(\frac{2\pi t}{T}\right), v = 1 + \cos\left(\frac{2\pi t}{T} + \pi\right)$, and $w = 1$. Here, time acts as an extra dimension. If the time-dependence is periodic with a period $T$, one can associate it with a wave number $\alpha = \frac{2\pi t}{T}$. Thus, an effective Chern number is defined by

$$C = \frac{1}{2\pi} \int_0^{2\pi} dk \int_0^{2\pi} d\alpha \, (\partial_k A_\alpha - \partial_\alpha A_k), \tag{6}$$

where $A_{k,\alpha} = i\langle\psi|\partial_{k,\alpha}|\psi\rangle$ is the Berry connection [14]. The value of the Chern number directly connects with the number of pumped particles: as time is adiabatically varied for a period, there are integer number of edge states crossing through the band gap (Fig. 2d), which is exactly the value of the Chern number. This quantized topological pumping can be seen as a manifestation of the bulk-boundary correspondence of the integer quantum Hall effect, also known as the Thouless pumping [63].

There has been a large amount of research efforts already in this area, among which one notable model is the AAH model (also called "Aubry-André (AA) model" or "Harper model" in some literature) [44, 45]. The AAH model describes a 1D chain where the on-site potential is spatially modulated, with the following Hamiltonian:

$$H = t\sum_N a_n^\dagger a_{n+1} + \lambda \cos(2\pi b n + \phi) a_n^\dagger a_n. \tag{7}$$

There is another off-diagonal version of the AAH model, featuring similar properties and being easier to implement practically (e.g., by using curved waveguides), and its Hamiltonian reads:

$$H = t \sum_N (1 + \lambda \cos(2\pi b n + \phi)) a_n^\dagger a_{n+1}. \quad (8)$$

In both Hamiltonians, $b$ characterizes a spatial modulation. A rational $b$ indicates a commensurate modulation, resulting in a periodic system, therefore being described by the Chern number ($\phi$ plays the role of $t$ in the previous discussion). However, when $b$ is irrational (an incommensurate modulation), the lattice is instead quasiperiodic. In this case, the Chern number can still be evaluated in real space, and the principle for quantized topological pumping is still valid [64, 65].

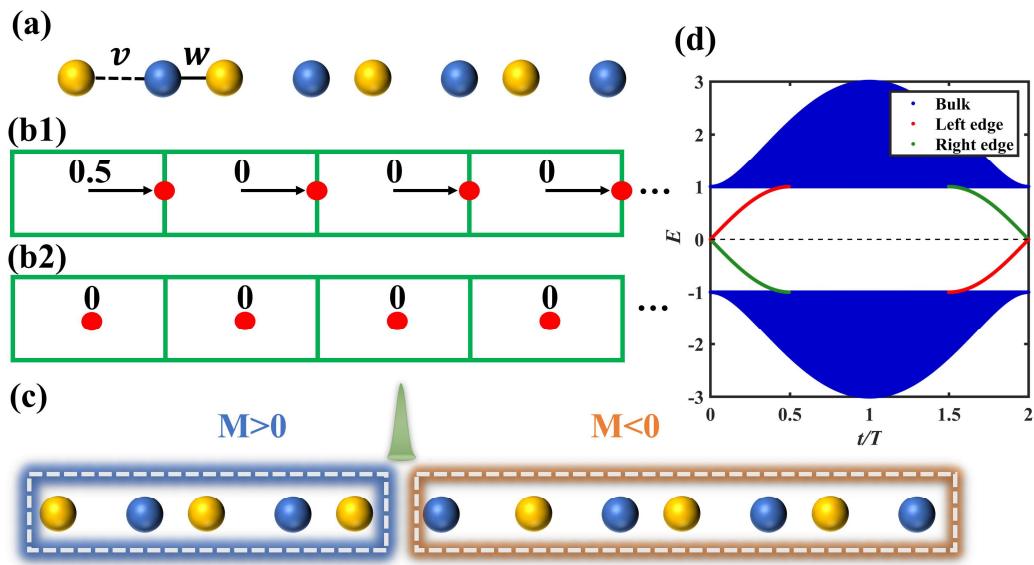

**Fig. 2 Illustration of the 1D topological model.** (a) Schematic diagram of the 1D SSH model, containing $A$ (yellow) and $B$ (blue) sublattices, where the nearest-neighbor coupling coefficients $v$ and $w$ are marked. (b1-b2) Illustration of the Wannier center (red dots) distribution for the topologically non-trivial (b1) and trivial (b2) regimes. The unit cells are denoted by green rectangles, wherein the filling charge is explicitly given. In (b1), there is a fractional (1/2) boundary charge, as a manifestation of non-trivial topology. (c) Schematic description of the Jackiw-Rebbi mechanism. At the interface of the two SSH chains possessing opposite Dirac masses, a bounded zero-energy mode emerges. (d) The spectrum of the Rice-Mele model as a function of $t$. Here $u = \sin\left(\frac{2\pi t}{T}\right), v = 1 + \cos\left(\frac{2\pi t}{T} + \pi\right), w = 1$, and the Chern number is 1. As $t$ is slowly varied from 0 to $T$, one left (right) edge state evolves into one right (left) edge state. The edge states are topologically connected with the Chern number, whose value indicates the number of pumped states from the left (right) boundary to the right (left) boundary.

Apart from the topological pumping in periodic optical structures or quasiperiodic crystals, the concept of topological pumping has also been extended to disordered and nonlinear systems, facilitating versatile developments in a variety of related fields [66-69].

# III. Topological photonics in 1D systems – Experiments

In this section, we explore the implementation of the SSH model across diverse photonic platforms, encompassing nanocavities, photonic crystals, and waveguide arrays. Our focus lies on elucidating our experimental methodologies for generating 1D photonic lattices, achieved through optical induction or continuous-wave (CW) laser-writing techniques in photorefractive strontium barium niobate (SBN) crystals. Furthermore, we delve into our recent contributions to nonlinear topological photonics, include realization of weakly nonlinear topological gap solitons, nonlinearity-induced nontrivial mode couplings, and the dynamics of topological phase transitions within interacting soliton lattices. Additionally, we explore the interplay of nonlinearity within the realm of non-Hermitian topological SSH lattices. Through these discussions, we hope to showcase the versatility of the SSH model across various photonic platforms and shed light on the intricate dynamics of nonlinear topological photonics.

## 3.1 The SSH model realized in various experimental platforms

Since the first realization of topological edge modes in photonic SSH lattices [26], the SSH model has been successfully implemented in a variety of experimental settings, including, photonic crystals (Fig. 3a) [70], waveguide arrays (Fig. 3b, c) [71, 72], microwave resonator chains (Fig. 3d) [73], coupled polariton micropillars (Fig. 3e) [74], Mie-resonant dielectric nano-disks (Fig. 3f) [75], and coupled fiber loops (Fig. 3g) [36]. The rich potential of the SSH model provides valuable insights into studying topological phenomena in optics, acoustics, condensed matter, and quantum physics. While several reviews have discussed the realization of topological models in different optical platforms [14-23], our focus here is mainly on implementing the SSH model using nonlinear photonic lattices. Photonic lattices, considered as a highly effective artificial photonic platform, offer a unique approach to studying discrete phenomena in optics. This effectiveness stems from the analogy between the paraxial equation for electromagnetic waves in optics and the Schrödinger equation in quantum mechanics. In essence, the evolution of a wave function over time in quantum systems is mathematically equivalent to the paraxial electromagnetic waves traveling in space [76, 77]. This equivalence provides a remarkable means to emulate the quantum dynamics of wave functions using classical electromagnetic waves in optics.

The optical induction technique in photorefractive crystals, utilizing the nonlinear photorefractive effect, was initially developed for the exploration of discrete phenomena such as optical spatial solitons [76, 77]. Unlike the permanent alterations of the refractive index encountered with the fs-laser-writing technique in materials like fused silica glass [78], optically induced photonic lattices in photorefractive crystals are reconfigurable and non-permanent [79, 80]. The index change in photorefractive crystals results from the intensity-dependent photorefractive effect, allowing the creation of different index potentials based on the intensity patterns from the interference of the induction beams. Despite the advantages, especially with the optical nonlinearity of photorefractive materials, the interference-based optical induction technique has limitations compared to the fs-laser-writing technique. For example, achieving a long-length waveguide or point-to-point writing is challenging due to the limited crystal length (typically 20 mm), thus hindering the observation of phenomena appreciable only over long distances of beam propagation. Complex lattices, such as helical waveguides in Floquet topological insulators or wiggled waveguide arrays for PT-symmetric lattices, are virtually impossible to achieve using conventional optical induction techniques.

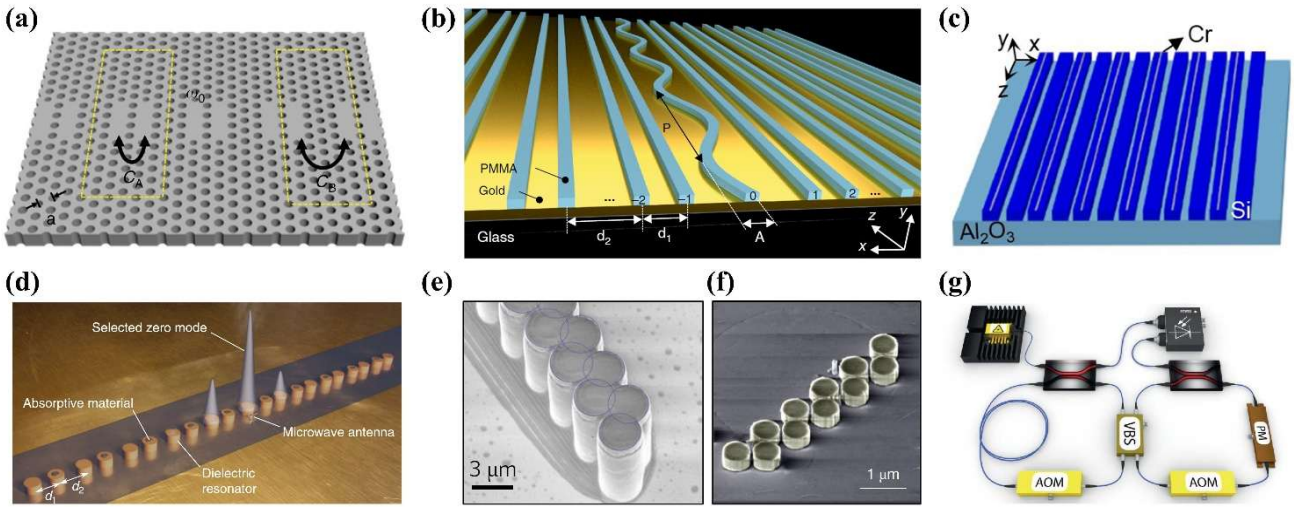

**Fig. 3 Photonic SSH lattices realized in various platforms.** (a) The nanocavity dimer array in the photonic crystal [70]. (b) The SSH lattice implemented in a plasmonic waveguide array [71]. (c) Non-Hermitian SSH model in silicon waveguide [72]. (d) Topological defect in the SSH model realized by a microwave resonator chain [73]. (e) The SSH chain of coupled polariton micropillars [74]. (f) The SSH array of Mie-resonant dielectric nanodisks [75]. (g) The SSH model realized with coupled fiber loops [36].

To address the limitations in constructing 2D lattices, we developed a method for CW-laser-writing of 1D SSH photonic lattices, as depicted in Fig. 4a. The method employs a collimated beam (wavelength 532nm, power 100mW) modulated by a spatial light modulator (SLM) serving as both the writing and probing beams. For the writing beam, the SLM encodes phase information, creating a stripe-shaped Gaussian beam pattern with adjustable waist and position. The beam is then spatially filtered by a 4F system, generating a stripe Gaussian beam that covers a (typically 20-mm-long) photorefractive crystal. To achieve a smaller lattice period, a set of cylindrical lenses compresses the beam width before entering the crystal. The waist of the writing beam is positioned inside the crystal for quasi-nondiffractive behavior while inducing refractive index changes. The memory effect of the photorefractive crystal ensures that the index changes remain intact for more than an hour. On the probing path, a Gaussian probe beam is filtered and collimated through controlled lenses, allowing precise manipulation of linear or nonlinear evolution in the lattice by switching the bias field. This approach can construct arbitrary 1D lattice structures with a spacing of around 15μm between nearest neighbor sites. The lateral illumination of the writing beam determines the waveguide length along the z-direction. This setup illustrates an effective approach for realizing SSH photonic lattices to explore linear and nonlinear topological phenomena, including trivial lattices, nontrivial lattices with edge and interface defects (Fig. 4b), and non-Hermitian SSH lattices as discussed in later sections.

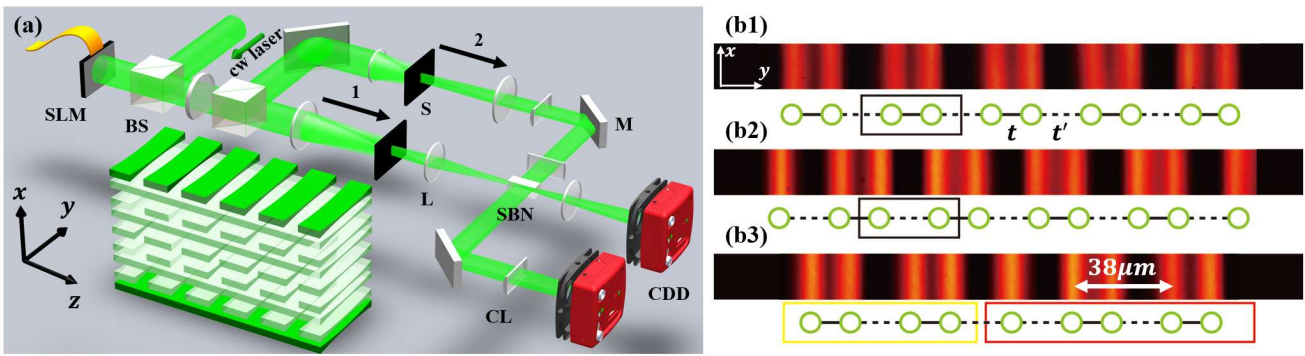

**Fig. 4 Experimental scheme for CW-laser-writing photonic lattices in photorefractive nonlinear crystals.** (a) Experimental setup for writing (laterally) and probing (longitudinally) a 1D photonic lattice. Path 2 is for the lattice-writing beam (ordinarily polarized), and Path 1 is for the probe beam (extraordinarily polarized). Lower inset shows the magnified lattice structure. (b) Illustration of different configurations of the SSH lattice. $t$ and $t'$ represent strong and weak coupling, respectively, and the black rectangle denotes a unit cell. (b1) and (b2) are the trivial and nontrivial SSH structures fabricated in the experiment, respectively. (b3) The SSH-type lattice with an interface ("long-long defect" formed with one trivial (squared by yellow lines) and one nontrivial (squared by red lines) dimer chains [29] (see also discussions in Section 2.1).

## 3.2 Nonlinear topological phenomena in SSH lattices

Nonlinear topological phenomena can be broadly categorized as "inherited" [29] and "emergent" [81] topological phenomena mediated by optical nonlinearity in the context of nonlinear topological photonics [19]. Emergent nonlinear effects manifest when an otherwise topologically trivial linear system becomes topological upon introducing nonlinearity [29, 30]. Notably, nonlinearity-induced topological phase transitions are a key manifestation of such emergent phenomena [82]. In nonlinear waveguide lattice structures [81], topological phase transitions occur when power surpasses a threshold value. Arrangements of interacting soliton lattices can also undergo dynamic emergent topological phase transitions [30]. In contrast, inherited nonlinear phenomena arise when nonlinearity is a small perturbation of a linear topological system. In 1D and 2D SSH lattices [33, 83], nonlinearity disrupts the chiral symmetry, breaking the underlying topology and enabling coupling to topologically protected boundary states [29, 83, 84]. However, many properties, including the features of nonlinear topological edge and corner states, are preserved as they are inherited from the original linear systems [29, 85-87]. The following discussion delves into a more detailed description of nonlinear topological phenomena in 1D SSH lattices.

The propagation of a linearly polarized optical beam in a Kerr nonlinear medium is governed by a nonlinear Schrödinger equation (NLSE) [79],

$$i\frac{\partial \psi}{\partial z} + \frac{1}{2k}\frac{\partial^2 \psi}{\partial x^2} + \gamma |\psi|^2 \psi = 0, \qquad (9)$$

where $\psi(x,z)$ is the electric field envelope, with $x$ denoting the transversal coordinate and $z$ the propagation distance, $\gamma$ is the strength of a Kerr-type nonlinearity, and $k$ is the wavenumber in the medium.

Firstly, we highlight the findings concerning nonlinear topological interface states under both self-focusing and self-defocusing nonlinearity, summarized in Fig. 5a. Using the SSH lattice structure similar to the one depicted in Fig. 4b3, we observe that when a low nonlinearity is activated, the output beam pattern still resembles the linear topological state (left panel in Fig. 5a3), indicating the formation of a topological gap soliton under a weak nonlinearity. However, under strong nonlinearity, achieved by increasing the power of the probe beam, the output pattern undergoes significant changes. Specifically, under strong self-focusing nonlinearity, the probe beam is confined to the center defect (Fig. 5a3), indicating the formation of a soliton with a propagation constant within the semi-infinite gap. Conversely, under high self-defocusing nonlinearity, the output pattern becomes highly delocalized, spreading into the bulk of the lattice, as shown in Fig. 5a3. These findings demonstrate that topological states in SSH lattices can sustain only a weak nonlinearity, losing their characteristic features under a strong nonlinearity, in agreement with the theoretical predictions [88].

Secondly, we present our findings on the nonlinearity-induced coupling of light into a topological defect, as a typical example of nonlinear inherited topological phenomena [29]. Unlike the previous excitation, where the topological modes were excited by a probe beam initially launched straight along the defect channel, here, a tilted broad beam traverses multiple lattice sites towards the defect (Fig. 5b1). In a nontrivial lattice, a tilted broad beam (with $k_x = 1.4\pi/a$, and a lattice constant $a = 38\mu m$) cannot couple into edge waveguides due to the topological protection (Fig. 5b1). Nonlinear conditions break the chiral symmetry, leading to the nonlinearity-induced coupling of light into topological edge states (Fig. 5b1). The underlying mechanism is analyzed using the nonlinear coupled mode theory [29], explaining the experimental observations. The theory outlines three stages of mode-coupling dynamics (Fig. 5b2). Initially, the probe beam, undergoing nonlinear propagation, cannot excite the linear topological edge mode. In the middle stage, as the beam approaches the defect, it perturbs the lattice structure, preventing the existence of the topological edge mode. In the third stage, part of the probe beam evolves into a nonlinearly localized mode, while the rest escapes into the bulk. Importantly, our theory shows that the nonlinear edge mode inherits properties from the topological edge state. This distinction between inherited and emergent nonlinear topological phenomena reveals that inherited states resemble modes of the underlying linear system (Fig. 5b3). They are affected by nonlinearity without altering the gap or changing the topological invariants. The nonlinear edge mode exemplifies the interplay of topology and nonlinearity, raising intriguing questions for further exploration.

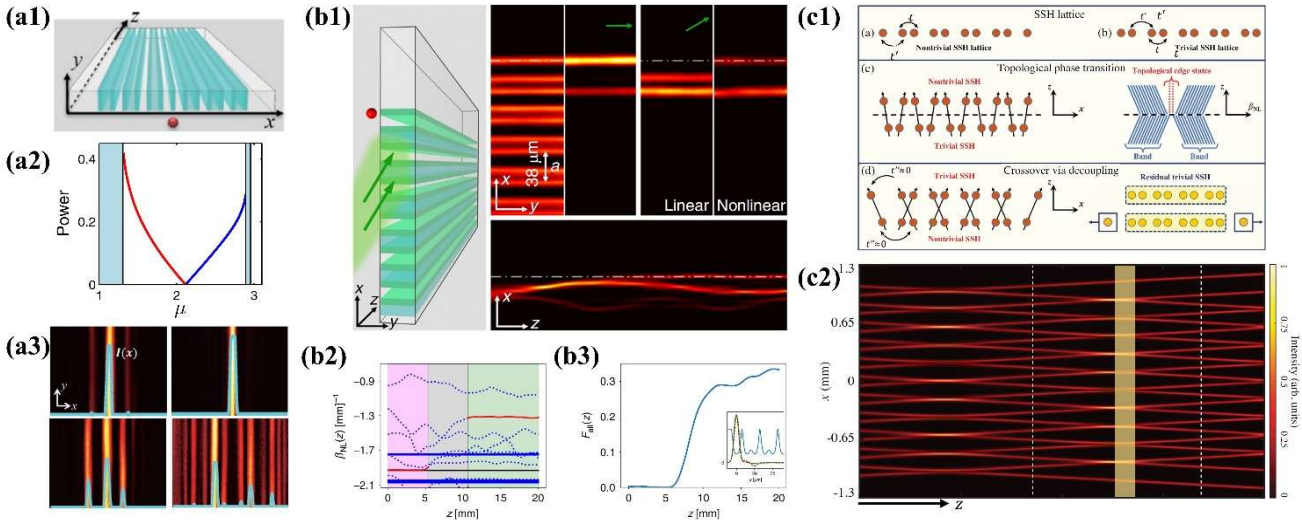

**Fig. 5 Examples of "inherited" and "emergent" nonlinear topological phenomena demonstrated in photonic SSH lattices.** (a) Nonlinear gap solitons observed in an SSH photonic lattice. (a1) A sketch of the SSH photonic lattice, where the red dot marks the interface "defect" waveguide. (a2) Power curves of solitons bifurcated from the linear topological interface mode of the SSH lattice under focusing (blue) and defocusing (red) nonlinearities. Shaded regions are Bloch bands. (a3) The top panel is the nonlinear output under weak (left) and strong (right) self-focusing nonlinearity. The bottom panel shows corresponding results under self-defocusing nonlinearity [88]. (b) Nonlinearity-induced nontrivial mode couplings. (b1) The left panel is an illustration of the excitation condition. The tilted beam aims at the edge but is launched from the bulk of a nontrivial SSH lattice. From left to right, at the top right panels are the lattice established in the experiment, linear output under normal (straight) excitation, and outputs with a tilted beam in linear and nonlinear cases. The bottom right panel shows a side-view (up to a crystal length of 20 mm) of the beam dynamics under nonlinear excitation in simulation. (b2) Nonlinear eigenvalue evolution under tilted excitation conditions (corresponding to the left panels of (b1)). The red (black) line denotes the nonlinear (linear) edge eigenvalue, while the individual blue dotted lines correspond to nonlinear localized states not inherited from the linear topological edge states. The three stages of the dynamics described in the text are denoted by the magenta, gray, and green-shaded regions. (b3) The overlap of the whole beam with the linear edge state to unveil the inherited topological properties of the excited states [29]. (c) The dynamic topological phase transitions in interacting soliton lattices. (c1) Top panel: the SSH lattice in topologically nontrivial and trivial regimes. Middle panel: sketch of the topological phase transition from trivial-to-nontrivial phase in real space (left) and in the spectrum (right). Bottom panel: sketch of the crossover from the nontrivial-to-trivial phase via decoupling of the outermost lattice sites. (c2) Intensity of the SSH soliton lattice evolving with propagation distance $z$ [30].

Apart from the inherited topological states discussed above, the dynamic topological phase transitions in the evolving nonlinear SSH soliton lattice can be seen as an emergent nonlinear topological phenomenon, existing only in the presence of nonlinearity. Figure 5c schematically illustrates the idea of such nonlinear dynamics [30]. As discussed earlier, a topologically nontrivial SSH lattice terminated with a weak coupling $t'$ at the end (Fig. 5c1) supports localized topological edge modes. In contrast, a trivial lattice terminated with a strong coupling $t$ at the end displays extended eigenmodes. From soliton interactions, a dynamic topological phase transition (from trivial-to-nontrivial phase) can occur along with a crossover from nontrivial-to-trivial regime (Fig. 5c1). Figure 5c2 presents the intensity distribution of the soliton lattice obtained from the NLSE calculations over a 16 mm-long propagation distance. Sublattices A and B collide periodically, maintaining their structures after each collision. The soliton lattice dynamically develops a topologically trivial SSH lattice until it reaches the distance marked by the first vertical dashed line. The spectral evolution can prove the topological phase transition, converting the lattice from trivial-to-nontrivial SSH geometry. Another topological phase transition occurs at the second dashed line (Fig. 5c2), leading to another pair of eigenmodes localized in the gap. However, in the region between these transitions, a crossover from the nontrivial to the trivial SSH phase takes place. The system briefly converts from nontrivial to trivial, and a new pair of edge states emerges after the second transition point, coexisting with decoupled walk-off solitons due to the crossover.

In this section, we have focused on only a few examples of "inherited" and "emergent" nonlinear topological phenomena using the SSH photonic lattices, including the formation of topological gap solitons, the nonlinearity-induced coupling of topological defect states, and the dynamical topological phase transitions arising from interacting soliton arrays. These examples shed light on the intricate interplay between topology and nonlinearity in the simple 1D topological settings, which may be further enriched by introducing non-Hermiticity, gauge fields, and Floquet modulations in higher-dimensional settings as well as in synthetic dimensions.

### 3.3 Nonlinear non-Hermitian SSH lattices

In the previous section, we provided an overview of our recent research based on Hermitian SSH lattices, in which no loss is introduced and the total energy of the system is conserved. Contrasting with the Hermitian systems, non-Hermitian systems involve energy exchange with the environment, resulting in complex spectra. Intriguingly, non-Hermitian systems that satisfy the parity-time (PT) symmetry can still exhibit real eigenvalues. Therefore, the exploration of PT-symmetry in optics over the past decade has emerged as a groundbreaking approach for manipulating light in non-Hermitian systems by modulating the interplay between gain and loss [89-93]. For instance, experimental observations have unveiled linear PT-symmetric topological states with real eigenvalues in fs-laser-written non-Hermitian SSH lattices [94], and the breakup and recovery of topological zero modes in fabricated silicon waveguide non-Hermitian SSH lattices [72]. In this section, we present our investigation on nonlinear non-Hermitian SSH lattices (NNH-SSH), representing the pioneering demonstration of nonlinear tuning of PT symmetry and non-Hermitian topological states [37].

Figure 6a depicts the non-Hermitian PT-symmetric SSH model, where the red, green, and blue dots represent the "gainy," "neutral," and "lossy" waveguides, respectively. The Hamiltonian for this PT-symmetric SSH model, within the tight-binding approximation, is expressed as follows:

$$\mathcal{H} = t \sum_{n \in N_G} a_{n-1}^\dagger a_n + t' \sum_{n \in N_G} a_{n+1}^\dagger a_n + t' a_1^\dagger a_0 + h.c. + \sum_{n \in N_G} (\beta^* a_n^\dagger a_n + \beta a_{n-1}^\dagger a_{n-1}) + \beta_0 a_0^\dagger a_0,$$

$$N_G = \{-1, -3, -5, \cdots\} \cup \{2, 4, 6, \cdots\}. \tag{10}$$

Here, $\beta = \alpha + i\gamma$, where $\alpha$ is the real part of the waveguide potential, and $\gamma$ represents the gain or loss. The parameters $t$ and $t'$ denote the strong and weak couplings, respectively, illustrated by different distances between adjacent waveguides in Fig. 6a. The creation operator on the *n*-th site is denoted as $a_n^\dagger$, and the set of integers $N_G$ corresponds to all the "gainy" sites marked as red channels in the left panel of Fig. 6c.

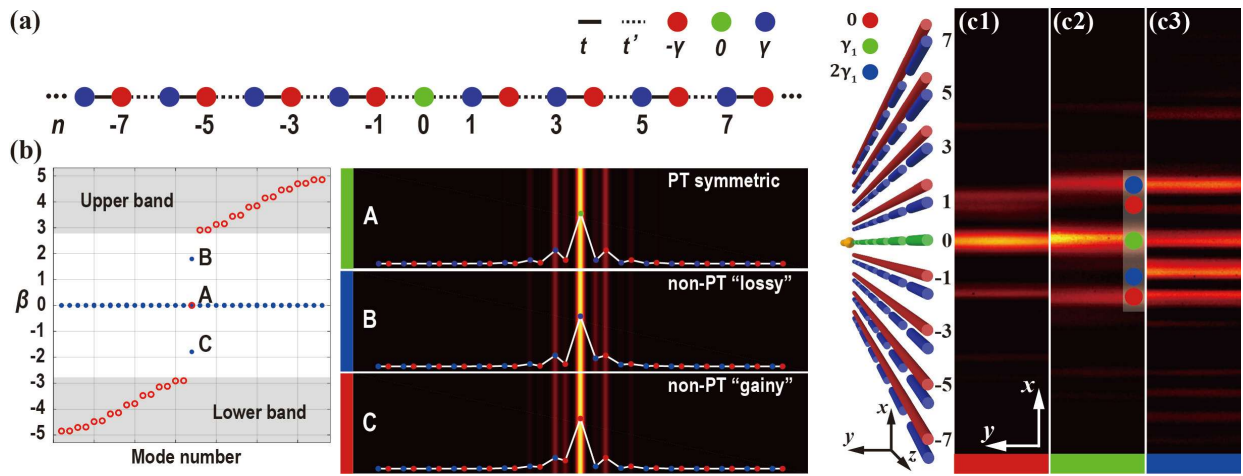

**Fig.6 Demonstration of nonlinear control of non-Hermitian topological edge states in SSH lattices.** (a) Illustration of an "active" PT-symmetric SSH with an interface topological defect located at site $n = 0$. Colored dots represent different lattice sites. (b) Left: Calculated eigenvalues for a finite lattice with 33 sites. Red circles and blue

dots denote the real and imaginary parts of the eigenvalues, respectively; shaded regions illustrate the band structure of an infinite lattice. Right: The corresponding eigenmode profiles, where the eigenvalues for points A to C are obtained with the propagation constants of the center waveguide $\beta_0 = 0$, $2i$, and $-2i$. Red, green, and blue dots represent "gainy", "neutral", and "lossy" lattice sites ($\gamma$ is the loss coefficient). (c1-c3) Experimental observations of nonlinear control of light localization in a defect waveguide in a "passive" PT-symmetric SSH lattice. On the left is a schematic of a lattice structure consisting of continuous and sectioned waveguides for a passive SSH lattice. (c2) The output, under linear excitation, exhibits the initial PT-symmetric feature of the topological mode, which is transformed by applying a self-focusing nonlinearity (c1) or a self-defocusing nonlinearity (c3). The red, green, and blue color bars indicate the system that is in a non-PT gainy, PT symmetric, and non-PT lossy regime, respectively. All results are adapted from Ref [37]

Conventionally, optical nonlinearity is perceived as an effect that alters the refractive index, potentially inducing a phase transition in PT-symmetric systems, as proposed in an early study [95]. Counterintuitively, as demonstrated in our recent work, despite photorefractive nonlinearity inducing a change solely in the real part of the refractive index, it concurrently affects both the real and imaginary parts of an induced waveguide potential. Consequently, once the passive PT-symmetric SSH lattice is established in our experiments, the optical nonlinearity can be employed to modify the imaginary part of the center waveguide potential, transforming the entire lattice into a non-PT-symmetric system. This transformation manifests in eigenmode profiles becoming asymmetric with respect to the center defect (Fig. 6b), serving as a distinctive signature of the phase transition between the PT and non-PT systems [37].

In demonstrating PT phase transition without actual gain in experimental systems, a global loss is typically introduced to convert the corresponding non-Hermitian PT-symmetric system into a passive PT-symmetric one [89, 94]. Our study employed the direct CW-laser-writing technique to establish non-Hermitian SSH lattices, introducing loss through sectioned waveguides (Fig. 6c). The gap size between sections allows for control over the amount of loss in each waveguide, creating "lossy" and "neutral" waveguides. Once the passive PT-symmetric SSH lattice is constructed, a stripe beam is sent into the center "neutral" waveguide channel. As the probe beam linearly traverses the lattice, a symmetric topological interface state proves the PT-symmetric feature [94]. Within the photorefractive crystal, self-focusing and defocusing nonlinearities are induced by applying a positive or negative bias field along the crystalline c-axis. Accumulating defocusing nonlinearity causes the probe beam to diverge from the center defect, leading to a highly asymmetric output (Fig. 6c2) and transforming the system into a non-PT "lossy" regime. Conversely, under self-focusing nonlinearity, the probe beam is directed more into the nearby "gainy" waveguide (Fig. 6c1), indicating a transition to a non-PT "gainy" regime. Notably, the sign of the nonlinearity in photorefractive crystals can be readily reversed [37], allowing reversible transitions between PT and non-PT regimes and enabling controlled destruction and restoration of PT-symmetric topological states.

This capability opens up many possibilities for studying nonlinear non-Hermitian topological phenomena, including, for example, exploring nonlinearity-induced birth and death of zero modes, topological gap solitons, and nonlinear control of exceptional points in non-Hermitian systems.

## 3.4 Topological photonics in 1D systems – Applications

In recent years, the exploration of the SSH model in the nonlinear regime has garnered significant attention for its potential applications in diverse photonic settings [18, 19]. In this section, we briefly discuss our recent work on the application of nonlinear SSH-type microstructures, shedding light on their capabilities in manipulating terahertz waves, inducing topological lasing in soft-matter superlattices, as well as achieving giant enhancements in nonlinear harmonic generation within nanocavity chains [96-98].

The growing interest in reliable terahertz (THz) technology is fueled by the increasing demand for applications like signal processing, biosensing, and non-destructive evaluation [99-103]. We first discuss our recent work on topologically tuned terahertz confinement in a nonlinear photonic chip. The SSH model in nonlinear terahertz (THz) photonics shows its potential for controlled THz generation and confinement in chip-scale devices [96]. The lattice, etched in the lithium niobate (LN) chip, features a one-dimensional SSH structure with waveguide stripes and wedge-shaped air gaps. The distances between neighboring LN stripes are given by: $d_1 = d_{10} - \delta d\, z/L$, $d_2 = d_{20} + \delta d\, z/L$, where $L = 6$ mm is the total length of the LN chip along the $z$-axis, and the distance $z$ is measured from the top of the chip (Fig. 7a), and the dimer structure at the top features $d_{10} = 80$ μm and $d_{20} = 30$ μm. As seen in Fig. 7a, the defect is located at the center ($n = 0$), but it varies from a long-long defect (L-LD) (when $z < L/2$) to a trivial equidistance without defect (at $z = L/2$), and then to a short-short defect (S-SD) (when $z > L/2$), thereby achieving different topological phases in these three different regions (illustrated by different colors in Fig. 7a). Thus, by using the femtosecond-laser pumping, topologically-tuned nonlinear generation and confinement of THz waves within the 0.1~0.8 THz frequency range is realized in the single LN chip, as the lattice undergoes topologically trivial to nontrivial transitions.

Secondly, robust topological interface state lasing within a designed polymer-cholesteric liquid crystal (P-CLC) superlattice is demonstrated based on the SSH model [97]. The chiral characteristics of the CLC's constituent molecules induce inversion symmetry breaking, favoring circularly polarized lasing. Experimental results, illustrated in Fig. 7b, demonstrate that right-handed circular polarization in the emitted laser light is achieved. Combining different P-CLC superlattices, the tunability of lasing wavelengths through temperature-induced shifts in the CLC photonic bandgap is enabled. Figure 7b illustrates the thermal tuning of lasing wavelengths, showcasing successive peaks at decreasing temperatures (581, 567, 556 to 544 nm). The high localization of topological states results in a low threshold for topological lasing (approximately 0.4 µJ), maintaining robustness even in the presence of substantial structural perturbations. The potential for strain tuning of the lasing wavelength on flexible substrates offers great potential for practical applications, including polarizers or color filters [104], intelligent responsive devices [105], and mirrorless lasing [106].

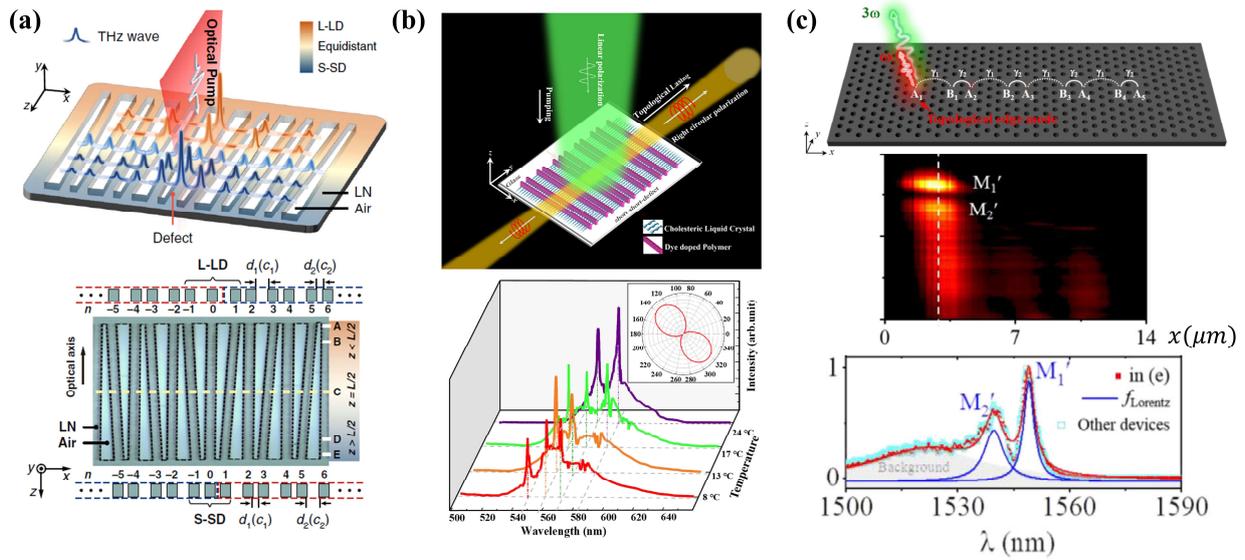

**Fig. 7 Examples of demonstrated applications of photonic SSH lattices under nonlinear processes.** (a) The nonlinear generation and confinement of THz waves in an SSH-type microstructure etched in a lithium niobate (LN) photonic chip, which undergoes a transition from long-long defect (L-LD) to short-short defect (S-SD) regions along $z$ (top panel). The microscope image of the LN array (50 µm thickness, $L = 6$ mm) with equidistant ($z = L/2$, $d_1 = d_2 = 55$ µm) and non-equidistant regions is shown in the bottom panel, where the yellow dashed line marks the L-LD to S-SD transition. A–E locations labeled on the right correspond to $z = 0$, $L/8$, $L/2$, $7L/8$, and $L$. Red/blue dashed lines indicate topologically nontrivial/trivial parts of the SSH lattice [96]. (b) Topological lasing in a polymer-cholesteric liquid crystal superlattice. The top panel shows the schematic of the TIS-lasing assembly with a linearly polarized green pump beam and right circularly polarized laser emission in opposite directions. The bottom panel shows the results of the thermal tuning of the topological lasing wavelength. The inset displays the intensity of quarter-waveplate-transformed laser radiation as a function of polarization angle in polar coordinates, where the polar angle represents the transmission angle of the polarizer, and the radius represents transmittance [97]. (c) Nonlinear harmonic generation in an SSH-type nanocavity chain. The top panel shows the schematic of the SSH structure, comprising nine coupled PCNCs with alternating weak and strong coupling strengths ($\gamma 1 < \gamma 2$) to support enhanced third-harmonic generation via the topological edge mode. The middle and bottom panels show the scattering spectra of the nontrivial SSH structure with white dashed lines indicating two successive peaks (M1' and M2') corresponding to the enhanced third-harmonic generation [98].

Finally, experimental observation of strongly enhanced third-harmonic generation in a silicon SSH structure comprising coupled photonic crystal nanocavities is demonstrated (Fig. 7c) [98]. This robust enhancement is attributed to the topological localization of an edge mode within the SSH chain. The edge mode not only inherits the resonant properties of individual photonic crystal nanocavities but also exhibits topological features with robustness extending well beyond a single cavity (Fig. 7c). The fabricated nontrivial SSH structure demonstrates a giant third-harmonic generation, comparable to that obtained from a single nanocavity, with both exhibiting a three-order-of-magnitude enhancement compared to a trivial SSH structure. The spatial distribution of the edge mode reveals the suppression of the decaying tail in the third-harmonic generation, attributed to the cubic relationship between its intensity and that of the edge mode (Fig. 7c). Therefore, the SSH structure of photonic crystal nanocavities holds promise as a platform for single-mode low-threshold topological lasing, which may find applications in new generation of topological lasers [107, 108] and chirality-selective routers for optical information [109, 110].

These findings collectively underscore the rich landscape of possibilities that nonlinear SSH models offer in the realm of optics and photonics. The ability to manipulate and control light in unconventional ways, as evidenced by these diverse applications, marks a significant stride forward in the pursuit of advanced and tunable optical functionalities by use of topological photonics. As research in this field continues to evolve and advance, the nonlinear SSH scheme may play a pivotal role in shaping the future landscape of photonic technologies.

## IV. Topological photonics in other 1D settings

In the preceding sections, our primary focus was on 1D photorefractive photonic lattices, which offer ease of fabrication and reconfigurability with low-power CW lasers. Within these SSH photonic lattices, the determination of topological edge states relies on the concept of trivial/nontrivial winding numbers. Nonetheless, the realm of 1D systems can go well beyond the standard SSH model using different platforms, holding various captivating topological phenomena, including, those observed in Floquet systems [111, 112], edge and bulk modes undergoing Thouless pumping [65-67] mode morphing in the synthetic mode dimension [113], perfect excitation using supersymmetry [114], non-Hermitian topology [115-118] and triple phase transitions [119, 120]. In this section, we briefly review some of the recent advances in other 1D settings. It is important to note that due to the rapid development of the research area, our discussion of these examples is subjective and not intended to be exhaustive.

Recent progress on Floquet topological phases has shed new light on time-dependent quantum systems, among which the 1D Floquet systems have attracted a great deal of attention [121, 122]. A 1D Floquet system can exhibit topological $\pi$ mode and zero mode illustrated in the quasi-energy spectrum. Experimentally, in periodically bent ultrathin metallic arrays of coupled corrugated waveguides that emulate the periodically driven SSH model, anomalous Floquet topological $\pi$ modes under proper frequencies were observed, propagating along the boundary of the arrays (Fig. 8a) [111]. Such Floquet systems provide effective experimental settings and valuable insights for the study of topological phases and nonlinear processes. Similar photonic simulators can serve as a testing ground for phenomena related to time-dependent 1D quantum phases, such as Thouless pumping and dynamical localization.

Thouless pumping provides an effective way to demonstrate higher-dimensional phenomena in 1D systems, for which edge pumping in 1D systems can feature a unidirectional edge transition equivalent to that in the corresponding 2D systems. These topological states and adiabatic pumping not only exist in periodic systems [63, 123] but were also proposed and demonstrated in quasicrystals (Fig. 8b) [65]. Indeed, in the pioneering work of Kraus et al. [65], topological boundary states and adiabatic pumping via topologically protected boundary states in an AAH photonic quasicrystal were observed. It has been revealed that quasicrystals possess nontrivial topological properties, uniquely associated with dimensions beyond their intrinsic 1D domain. Notably, theoretical and experimental investigations have demonstrated that the 1D quasicrystals can be characterized by Chern numbers, featuring topologically protected boundary states analogous to those observed in a 2D quantum Hall system [65]. These findings, based on the 1D settings, offer a broader understanding applicable to higher-dimensional systems with diverse topological indices. Furthermore, the bulk pumping reveals the topological invariant of the system and is manifested as a quantized transverse displacement [66, 124]. When nonlinearity is considered, the concept of quantized nonlinear Thouless pumping can be introduced to show the topologically quantized transport of photons with a non-uniformly occupied band (Fig. 8c) [67]. Unlike traditional Thouless pumping, where quantization relies on uniformly filled bands, nonlinearity, and interparticle interactions can induce quantized transport through soliton formation and spontaneous symmetry-breaking bifurcations. The unique mechanism presents a departure from the conventional approaches, where instantaneous soliton solutions remain identical after each pumping cycle.

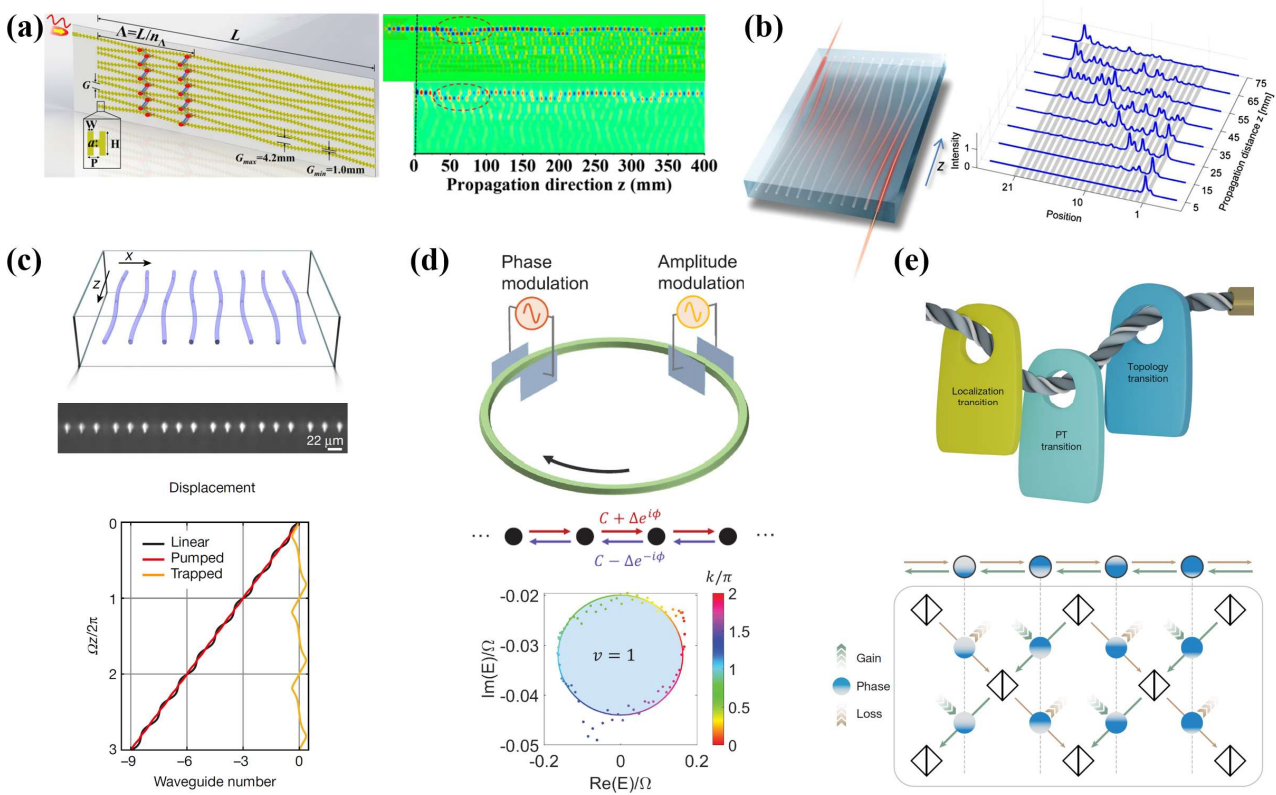

**Fig. 8 Examples of topological phenomena demonstrated in other 1D photonic settings.** (a) Observation of anomalous $\pi$ modes by photonic Floquet engineering. Left panel: Schematic of periodically bent ultrathin metallic waveguides with cosine modulation of lattice spacing G. Inset: an "H"-shape structure. Right panel: simulations (top) and near-field measurements (bottom) under various driving conditions for curved waveguides [111]. (b) Demonstration of topological states and adiabatic pumping in quasicrystals. Left panel: Illustration of slowly varying spacing between waveguides along the z-axis, creating an adiabatically modulated Hamiltonian for injected light. Right panel: The experimental results, when light is injected into the rightmost waveguide (site 1). Intensity distributions at different stages of adiabatic evolution (various propagation distances) show the light crossing the lattice from right to left [65]. (c) Observation of quantized nonlinear Thouless pumping. Top panel: 1D waveguide array with in-plane modulation. Micrograph of the output facet, displaying the lattice implementation. Bottom panel: Displacement of the center of mass during linear and nonlinear pumping processes [67]. (d) Demonstration of arbitrary topological windings of a non-Hermitian band. Top panel: Realization of non-Hermitian 1D lattices using a ring resonator undergoing simultaneous amplitude and phase modulation at integer multiples of the free-spectral range. Middle panel: Non-Hermitian 1D lattice with nearest-neighbor coupling. Bottom panel: The dots represent the band energy extracted from signals captured in the experiment. The lines denote theoretical predictions of corresponding band energy winding in the complex plane as $k$ ranges from 0 to $2\pi$ [118]. (e) Topological triple phase transition in non-Hermitian Floquet quasicrystals. Top panel: Intertwinement of a triple phase transition in the non-Hermitian Floquet AAH model. Combining a 1D quasicrystal with non-Hermitian anisotropy through temporal driving results in the Floquet AAH model. Bottom panel: Implementation of the 1D lattice with a quantum walk using a mesh lattice of beam splitters created with coupled fiber loops. The combined modulations of gain/loss and phase implement the non-Hermitian Floquet AAH model based on a discrete-time quantum walk [120].

The interplay between topology and non-Hermiticity has also turned into an intriguing subject of research in non-Hermitian systems [62]. A distinctive aspect of non-Hermitian systems is the nontrivial winding of the energy band in the complex energy plane [115, 116]. Experimental demonstrations of such nontrivial winding by implementing non-Hermitian lattices along a frequency synthetic dimension in a ring resonator have been realized with simultaneous phase and amplitude modulations (Fig. 8d) [118]. Complex band structures are directly characterized, highlighting the controllable nature of topological winding through modulation waveform changes. Those results show the ability to synthesize and characterize topologically nontrivial phases in nonconservative systems. Since the fiber loops can bring unprecedented flexibility to tailor the complex band structure, they have been employed to achieve non-Hermitian, long-range, and complex-valued couplings [117, 125, 126]. For example, the time pulse arrays in fiber loops are also utilized to construct the non-Hermitian Floquet quasicrystals, which can be employed to study a new type of phase transition [119]. Indeed, despite seemingly disparate concepts, it has been shown that bulk conductivity, topology, and non-Hermitian symmetry breaking are interconnected in a non-Hermitian quasicrystal system, demonstrating the triple phase transition in a 1D non-Hermitian synthetic quasicrystal implemented through a Floquet photonic quantum walk in coupled optical fiber loops (Fig. 8e) [120]. The results highlight the intertwinement of topology, symmetry breaking, and mobility phase transitions in non-Hermitian quasicrystalline synthetic systems.

The exploration of topological photonics in various 1D settings extends beyond the conventional SSH model, encompassing phenomena such as those observed in Floquet systems, Thouless pumping, and non-Hermitian synthetic quasicrystals. These studies not only enhance our understanding of topological phases and nonlinear processes but also provide valuable insights applicable to higher-dimensional contexts.

The significance of these investigations lies in uncovering unique topological features, such as Floquet topological modes, nontrivial edge states in quasicrystals, and the interplay between topology and non-Hermiticity. Such findings greatly contribute to the broader field of topological photonics, offering new possibilities for manipulating light in unconventional ways and potentially leading to the development of novel photonic devices with advanced functionalities [18, 19, 127].

## V. From SSH to higher-order topological insulators (HOTIs)

In recent years, there has been a great deal of interest in the study of higher-order topological insulators (HOTIs) – a class of topological insulators that support protected gapless boundary modes at boundaries of higher dimensions, e.g., the corners of a 2D material, broadening the concept of symmetry-protected topological (SPT) phases [19, 128-130]. HOTIs have been demonstrated in a variety of fields ranging from condensed matter physics to electric circuits, spintronics, mechanics, acoustics, and photonics [131, 132]. Compared to standard topological insulators that commonly obey the bulk-boundary correspondence principle, this new class of topological insulators does not respect the conventional bulk-boundary correspondence. A zero-dimensional corner state in 2D or higher-dimensional structures serves as a typical manifestation of the HOTIs regardless of physical dimensions. Fundamental studies and applications of HOTI corner states have been reported in diverse physical platforms in both linear [129, 133-135] and nonlinear regimes [83, 133, 136-139], as well as with orbital degrees of freedom [140].

One typical class of HOTIs is called the topological crystalline insulators (TCIs) [20, 129]. Many of the topological features characterized by filling anomaly and charge fractionalization in TCIs can be illustrated again from the simplest SSH model [53], represented by $C_4$-symmetric 2D SSH-type TCIs (protected by chiral symmetry), and $C_3$-symmetric Kagome-type TCIs (protected by generalized chiral symmetry). As a simple model for studying the physical mechanisms of corner states, the 2D SSH model [141, 142] can support linear higher-order corner-localized modes found in the bulk continuum (i.e., the bound states in the continuum, BICs) [143]. A 2D SSH lattice is composed of unit cells formed by four sites and satisfies chiral symmetry. It exhibits two distinct Zak phases accounting for the two extending directions, whose values can be tuned by varying the dimerization parameter. In the topologically nontrivial phase, where the intercell hopping is weaker than the intracell hopping, this system supports four degenerate BIC corner states embedded in the bulk band at zero-energy. Topological protection of corner states at zero-energy is guaranteed by chiral symmetry and $C_{4v}$ symmetry [144].

In this section, we briefly discuss our recent work on these two types of TCIs. The first type has been studied previously in the linear regime [143]. In the presence of nonlinearity, the BIC corner states of the 2D SSH lattice behave differently when compared to the linear regime, leading to nonlinear control of photonic higher-order topological BICs [83]. As schematically illustrated in Fig. 9a, corner states in a 2D topologically nontrivial SSH lattice can be excited at a single corner site under both self-focusing and defocusing nonlinear regimes. In this study, Hu et al. showed that nonlinearity plays a crucial role in determining the corner-state dynamics of the 2D SSH lattice for different dimerization conditions [83]. Remarkably, the application of a weaker nonlinearity induces not only the separation of corner states from the bulk band at zero-energy in the topological nontrivial phase but also transitions between corner and in-gap edge states under both self-focusing and defocusing regimes. Although the amplitude of nonlinear corner modes differs from the linear counterpart, the feature of topological states (corner localization with non-zero amplitudes only in one sublattice and exponentially decays in the bulk) is still evident, as it is "inherited" from the linear 2D SSH lattice. However, at high values of nonlinearity, the corner states are liberated from the continuum and move out of the Bloch bands to form semi-infinite gap solitons in the self-focusing regime, or undergo strong radiation into the bulk and edges in the self-defocusing case, no longer preserving the topological feature.

The second paradigmatic class of HOTI models that has received particular attention is the Kagome lattice protected by the generalized chiral symmetry [145]. Bulk polarizations with nonzero or zero quantized values are the topological invariants that characterize the topologically nontrivial or trivial phases of the breathing Kagome lattices (BKLs), where the appellative "breathing" intends that intercell and intracell hopping are not equal. The nontrivial (trivial) phase of BKLs takes place when intercell hopping is larger (smaller) than the intracell one. Two different types of BKL structures can be generally engendered by selecting either an upward- or a downward-unit cell, each of which is derived from distinct primitive generators of the $C_3$-symmetric TCI [53], resulting in BKLs that display a flat edge (Fig. 9b) or a bearded edge (see Fig. 9d) boundary. A nontrivial BKL with flat-edge truncation exhibits both singular band-touching and HOTI properties. In this regard, different finite-sized geometrical structures of nontrivial BKL are employed to investigate HOTIs in different physical systems as well as in both linear and nonlinear regimes [139, 145-149], with major emphasis given to the triangle (Fig. 9b) and rhombic (Fig. 9c) shapes. A triangle BKL features $C_3$-rotational symmetry and supports three topologically corner states located at zero-energy with characteristic amplitude and phase distributions in each sublattice (Fig. 9b), while the rhombic counterpart possesses only one corner state (e.g., the A sublattice in Fig. 9c).

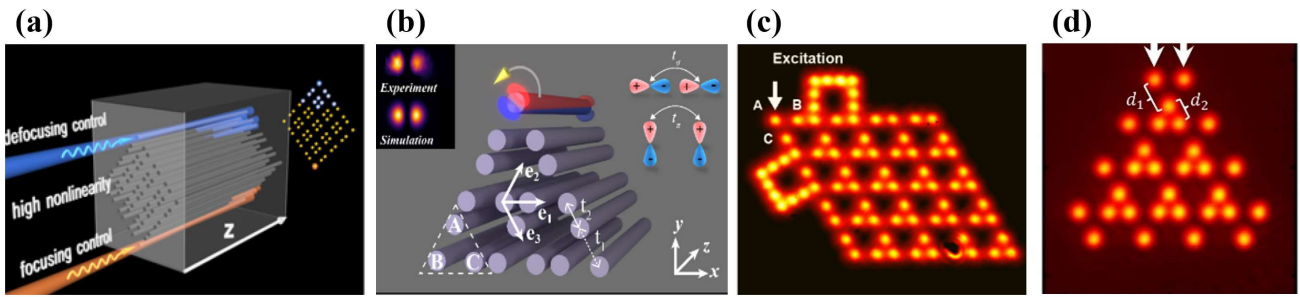

**Fig. 9**. **Examples of photonic topological crystalline insulators linked to the SSH lattices**. (a) Illustration of nonlinear control of photonic higher-order topological bound states in the continuum in a 2D SSH photonic lattice [83]. (b) Schematic illustration of a triangle BKL with $\sigma$- and $\pi$-type orbital hopping. Inset in (b) shows one $p$-orbital mode from experiment and simulation. Nonlinear control of a $p$-orbital corner state is illustrated by a yellow arrow [140]. (c) Experimental rhombic photonic BKL with nontrivial geometry and SubSy-breaking configuration, where A, B, and C label three sublattices at the corner unit cell, and the white arrow indicates the onset excitation. In the illustration, the SubSy for A sublattice is broken by two bridge waveguides introducing A-A coupling, affecting the topological protection of the corner state [61]. (d) Experimental photonic triangle BKL with bearded edge truncation in nontrivial geometry to observe in-phase and out-of-phase topological corner states. White arrows indicate corner state excitation at the two uppermost lattice sites [150].

Despite a vast amount of research efforts to demonstrate HOTI modes, most of these studies have been devoted only to topological corner states in the lowest $s$-band, with very limited theoretical studies focused on the orbital degree of freedom [151, 152]. Triangle-shaped BKLs have been recently adopted as a perfect candidate for experimentally exploring $p$-band orbital corner states and unveiling their underlying topological invariant and symmetry protection. Figure 9b shows that both $p_x$- and $p_y$-type orbital corner states in the nontrivial BKL possess the characteristic intensity and phase structures of zero-dimensional zero-energy modes, similar to those in the $s$-band. Nontrivial topology in the $p$-band of the BKL can be manifested by introducing the generalized winding number since bulk polarization suffers the bulk-band hybridization effect, leading to misinterpretation of the real effective values. Topological protection of $p$-orbital corner states require more restricted conditions than $s$-orbital corner states in the same system, since the orbital-hopping symmetry (e.g., $t_{1\sigma}/t_{2\sigma} = t_{1\pi}/t_{2\pi}$) must also be satisfied apart from the generalized chiral symmetry and $C_3$-rotational symmetry [140]. Notably, orbital-hopping symmetry is not required in the $p$-band 2D SSH model to protect orbital corner states, because the $p_x$ and $p_y$ subspaces are inherently decoupled [153]. Additionally, an interesting phenomenon taking place in the nonlinear regime of the nontrivial BKL is schematically illustrated in Fig. 9b, where a single-site corner excitation leads to the dynamical rotation of a dipole-shaped beam due to the nonlinearity-induced lifting of $p_x$- and $p_y$-orbital mode degeneracy.

Furthermore, a rhombic-shaped BKL with a single HOTI corner state has been employed to theoretically and experimentally demonstrate that boundary states are protected by pertinent sub-symmetry (SubSy) [61]. Generally speaking, an SPT phase is associated with a topological invariant characterizing the bulk and manifests symmetry-protected boundary states. Any perturbation preserving the protecting global symmetries was believed to constitute the necessary condition to retain the boundary states or the topological invariant in the system when the gap between bands remains open [14, 49]. The concept of SubSy raises a challenge to previous understandings of the SPT phases by demonstrating that topological boundary states can also exist even when the global symmetry is affected by appropriate perturbations that preserve the SubSy. In this case, under the SubSy-preserving perturbations, the boundary state remains intact and its eigenvalue is pinned at zero-energy, even though the overall topological invariant is lost. In this work, by establishing 2D photonic rhombic

BKLs with long-range coupling introduced by two bridge waveguides (Fig. 9c), the protection of corner states at the A site is tested. The bridge waveguides are appropriately positioned in the BKL structure to introduce perturbations in different sublattices that break or preserve the associated SubSy. Experimental results prove that, in A-SubSy preserving perturbations, the corner state is robust and light remains confined to the A-sublattice sites after propagation, whereas in the A-SubSy breaking lattice, a corner excitation leads to coupling with other sublattice sites. Moreover, by adding non-negligible long-range hopping to BKLs, the existence of a long-range hopping symmetry is uncovered, which is essential for the protection of the corner states [61].

Currently, much of the exploration of HOTI corner states has focused on the flat-edge BKL structure, but not much attention has been devoted to its bearded-edge version. Motivations are ascribed to the fact that topological corner states in bearded-edge BKLs are not HOTI states per se, since the fractional corner anomaly is zero in this model [154]. Note that the fractional corner anomaly is the most used topological invariant for discriminating HOTI properties in real space. However, a bearded-edge BKL can support two distinct classes of topological corner states with in-phase and out-of-phase structures [154]. Specifically, the in-phase corner states (IPCSs) only appear in the topologically nontrivial phase, highly localized at the two outermost sites with the same phase relation and the nonzero amplitudes in more than one sublattice. In contrast, out-of-phase corner states (OPCSs) manifest in both topologically trivial and nontrivial phases, either as BICs or in-gap states, depending on the lattice dimerization conditions, and their amplitude distribution is strictly localized to the two outermost sites with an opposite phase. In this context, Song et al. in Ref [150] reported the first experimental demonstration of IPCSs and OPCSs, achieved in a 2D photonic triangle bearded-edge BKL with different trivial and nontrivial geometries (Fig. 9d). More importantly, they disclosed that these two types of corner states feature different topological origins: IPCSs are protected by momentum-space topology and characterized by nonzero bulk polarizations. Instead, OPCSs are protected by both momentum- and real-space topology, associated with strongly localized flat-band states in any lattice dimerization. Both share topological protection from the $C_3$-rotational and generalized chiral symmetries, exhibiting nonzero-energy mode features, but the OPCSs additionally require protection from real-space topology to preserve distinctive characteristics of perfect eigenvalue degeneracy, with compact amplitude localization at two corner sites solely.

In summary, this section has presented an overview of our recent work on 2D topological photonic lattices based on the SSH model. These lattices exhibit intriguing linear and nonlinear phenomena associated with topological corner states. Our findings contribute to the understanding of HOTIs and hold potential for the development of novel photonic devices. It may prove useful to the subsequent development of novel generations of photonic devices.

# VI. Conclusion and outlook

The chapter provides an extensive overview of recent advancements in topological photonics within 1D systems, with a specific emphasis on photorefractive photonic lattices. It covers the realization and manipulation of both linear and nonlinear topological edge states within the context of the SSH model. Furthermore, the exploration extends beyond the SSH model to encompass various other 1D settings, such as Floquet engineering, Thouless pumping, and non-Hermiticity, thereby highlighting the breadth and diversity of current research in this field. In particular, experimental progress based on photorefractive lattices and photonic crystals is reviewed, showcasing the versatility and effectiveness of these platforms in studying topological phenomena. The chapter also discusses the role of optical nonlinearity in facilitating the study of nonlinear topological photonics, including topics such as topological edge states, non-Hermitian SSH lattices, nonlinear tuning of topological states, and topological lasing. Furthermore, the utilization of topological pumping and HOTIs is discussed, illustrating the potential for manipulating light in novel ways. These advancements not only contribute to our fundamental understanding of topological physics but also hold promise for enhancing the robustness and functionality of various photonic devices, including resonators, delay-line devices, waveguide arrays, and lasers (Fig. 10).

Indeed, research on 1D topological systems has played a central role in advancing the broader field of topological photonics by uncovering a rich variety of physical phenomena and enabling the development of novel photonic devices with advanced functionalities. Some of the key contributions and implications include:

- Exploration of novel phenomena: Studies of 1D topological systems have revealed a diverse range of phenomena, such as Floquet topological modes [112, 155], topological pumping [68, 69, 156], synthetic multi-dimensionality [40, 157], the interplay between topology and non-Hermiticity [115, 127, 158, 159], dissipative coupling [39, 160], and quantum interference [161–163]. These discoveries deepen our understanding of topological physics and expand the theoretical framework for describing light–matter interactions in nontrivial structured environments.

- Technological applications: The insights gained from research on 1D topological systems pave the way for the development of innovative photonic devices with advanced functionalities. For example, the realization of topological lasing [18, 19, 28, 70, 74, 164], enabled by the robustness of topological edge states, holds promise for applications in telecommunications, sensing, and quantum information processing.

- Advancement of fundamental science: Beyond technological applications, research on 1D topological systems contributes to fundamental science by uncovering new principles governing the interaction of light with structured materials such as photonic crystals, waveguide lattices, and metamaterials. These insights have implications for diverse areas of physics and materials science, enriching our understanding of complex quantum systems and condensed-matter phenomena.

To update briefly the progresses over the recent two years, we notice that 1D topological photonics has evolved from canonical SSH-type realizations toward a broader and more functional framework, emphasizing symmetry flexibility, programmability, disorder, and active control. Extending the Shockley-like surface states first observed in photonic superlattices [26], recent studies have demonstrated that robust boundary localization can persist under relaxed or partial symmetry constraints, including subsymmetry-protected phases and topological bound states in the continuum [165]. In particular, subsymmetry has enabled the construction of topological bound states in the continuum in effectively one-dimensional photonic lattices, generalizing the bulk–boundary correspondence beyond conventional chiral-symmetric models [165]. At the same time, disorder has emerged as a constructive resource rather than a detrimental factor: both gapped and ungapped photonic topological Anderson insulators have been experimentally realized, showing that disorder can induce, rather than destroy, topological phases in 1D systems [166].

A defining advance during this period is the rapid development of programmable and reconfigurable integrated photonic platforms, which allow dynamic implementation and continuous tuning of 1D topological Hamiltonians on a single chip [167]. These systems enable real-time switching between trivial and topological regimes, controlled creation and displacement of interfaces, and systematic exploration of parameter space well beyond static superlattice designs. In parallel, new physical realizations of 1D topology have been reported, including topological helix chains, synthetic-dimension lattices, and Floquet-driven systems, where anomalous and switchable edge states emerge without static counterparts [168].

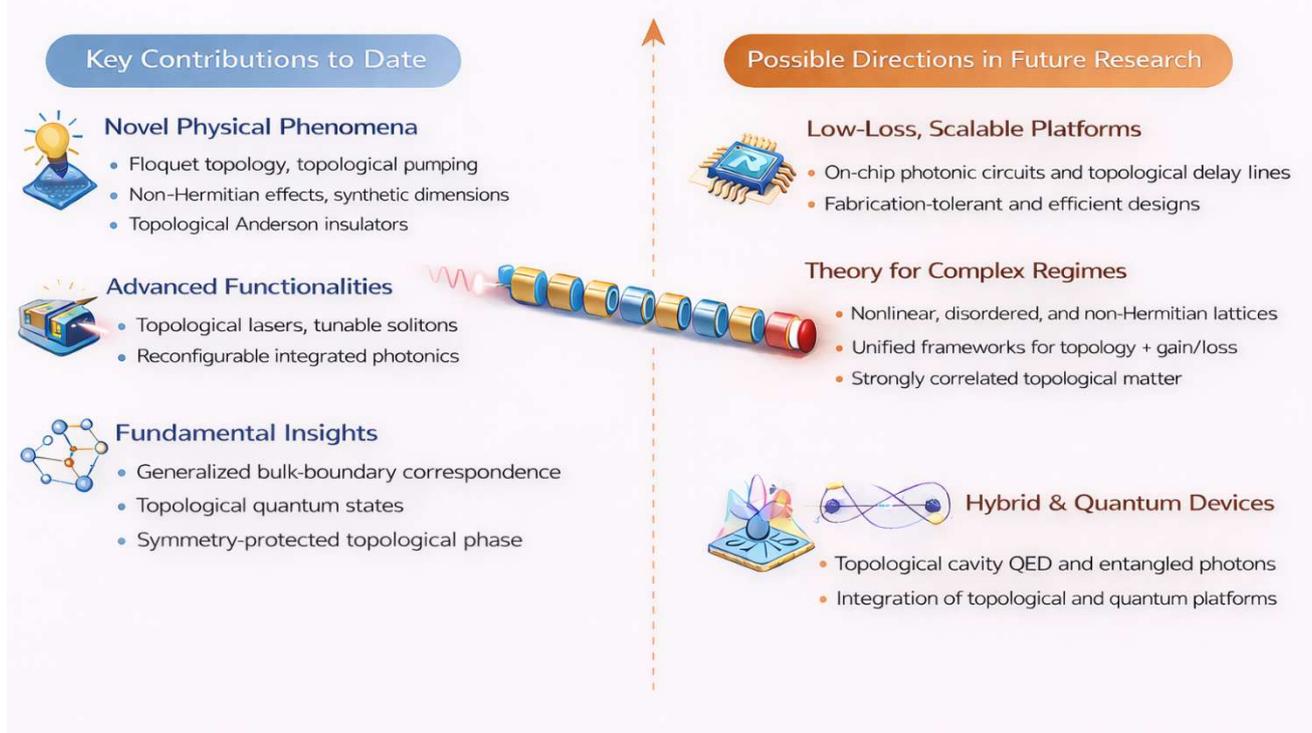

Fig. 10. **List of some key contributions and possible future directions of 1D topological photonics.**

Beyond existence proofs, recent efforts increasingly focus on functionality and applications. Non-Hermitian photonic lattices now support tunable edge states and topological solitons with enhanced sensitivity and controllable response [169–173], while engineered interface states in metallic and terahertz waveguides demonstrate broadband signal-processing capabilities and robustness against structural imperfections [174]. These application-oriented directions are accompanied by a deeper theoretical understanding of non-Hermitian bulk–boundary correspondence and response functions in one-dimensional systems [175]. Recent reviews further contextualize these advances, highlighting both the opportunities and fundamental limitations of topological photonics [176, 177], as well as emerging space–time and quantum-walk perspectives that enrich the conceptual landscape of 1D topology [178].

Looking ahead, several important challenges and open questions remain for one-dimensional topological photonics. A central issue is how to achieve scalable, low-loss, and fabrication-tolerant implementations of topological lattices while preserving symmetry control and mode selectivity in realistic platforms. The interplay between topology, disorder, nonlinearity, and non-Hermiticity— particularly beyond perturbative regimes—remains incompletely understood and calls for unified theoretical frameworks capable of capturing nonlinear, dynamical, and dissipative effects on equal footing. Another open challenge lies in extending 1D topological concepts toward functional quantum and hybrid photonic systems, where coherence, gain, and noise compete, and where topological protection must be reconciled with strong light–matter interactions. From an application perspective, translating robust boundary modes into practical device architectures, such as reconfigurable delay lines, topological lasers, and on-chip signal processors, requires systematic studies of bandwidth, efficiency, and stability under real-world operating conditions. Addressing these challenges will be crucial for transforming 1D topological photonics from a primarily exploratory framework into a mature and impactful platform for both fundamental science and photonic technologies.

Overall, we envision that research on 1D topological systems will continue to play a pivotal role in shaping the future of topological photonics and light-field manipulation, bridging fundamental discoveries and practical device concepts, and opening new avenues for scientific exploration and technological innovation.

**Acknowledgment:**

We thank D. Leykam, D. Smirnova, K. Makris, A. Szameit, J. Yang, X. Zhang, X. Gan, J. Zhao, and all our co-workers for collaboration and assistance on related work presented in this review.

**Funding:** This research is supported by the National Key R&D Program of China under Grant No. 2022YFA1404800, the National Natural Science Foundation (W2541003, 12134006, 12274242, 91750204, 11674180, 12474387, 12250410236), PCSIRT, and the 111 Project (No. B07013) in China. H.B. acknowledges support from the project "Implementation of cutting-edge research and its application as part of the Scientific Center of Excellence for Quantum and Complex Systems, and Representations of Lie Algebras, Grant No. PK.1.1.10.0004, co-financed by the European Union through the European Regional Development Fund－Competitiveness and Cohesion Programme 2021-2027. D.B. acknowledges support from the 66 Postdoctoral Science Grant of China and the Ministry of Human Resources and Social Security of China (Grant WGXZ2023110).

**Disclosures.** The authors declare no conflicts of interest.

**Authors' contributions**:    All authors contributed to this work.